\documentclass[prb,twocolumn,showpacs,floats,eqsecnum,amsmath,amssymb]{revtex4}
\usepackage[dvips]{graphicx}

\begin{document}


\title {Zero vorticity condition in calculation of ground state energy of Josephson junction lattices}

\author {Y. Azizi}

\affiliation{ Institute for Advanced Studies in Basic Sciences,
Zanjan 45195-1159, Iran}

\begin{abstract}
We employ the charge neutrality condition in the form of zero
vorticity for a Josephson junction lattice in calculating its
ground state. We consider a Fibonacci ladder and the Penrose
lattice to test our method. We also compare the results with
those of the model that treats the plaquettes independently. We
show that the zero vorticity condition improves on the results of
the independent plaquette model.
\end{abstract}

 \maketitle
\section{Introduction}
A Josephson junction is characterized by its current-phase
relation $I(\phi)$, where $\phi$ is the phase difference across
the junction. $I(\phi)$ must have two general properties: it must
be $2\pi$- periodic and it must be an odd function. Therefore
$I(\phi)$ can be any periodic, odd function
\cite{likharev,likharev2,GK}. In here we focus on sinusoidal
Josephson junctions. In this case, the Hamiltonian of system is a
cosine function of the gauge invariant phase difference
\cite{GK,Tinkham}. A Josephson junction lattice is a lattice with
nodes that are superconducting islands and bonds that represent
Josephson junctions. This lattice of Josephson junctions is a
model that can be used for the description of other important
systems like stacked Josephson junctions \cite{CG},
high-temperature superconductors \cite{MFF,MFF1,MFF2}, and a
rotating BEC \cite{kk}.

If we ignore screening effects, we can write the Hamiltonian of
the system as a sum of Hamiltonians of Josephson junctions
\cite{Tinkham}. In here we focus on minimization of this
Hamiltonian as a function of superconducting phase in presence of
external magnetic field \cite{k2002}. By this minimization we
obtain the ground state energy as a function of the external
magnetic field, or frustration factor, which is defined as the
ratio of flux through a plaquette with unit area to the quantum
of flux. Behavior of ground state energy $E$ as a function of
frustration factor $f$ is interesting. When $f$ is zero all
junctions can attain their absolute minimum, but when $f$ becomes
nonzero, the bonds of the lattice cannot generally obtain this
absolute minimum simultaneously and produce some superconducting
current--this is the origin of the term {\it frustration}.

We approximate the behavior of the lattice in ground state by the
Independent Plaquette Model (IPM), \cite{k2002} which considers
the  plaquettes independently. Also we suppose that total
vorticity of lattice is zero. This condition can be understood
once we look at the relation of this model with Charged Coulomb
Gas (CCG) model. The Hamiltonian of the Josephson junction
lattice can be mapped to CCG on dual lattice, where the charge of
each point in dual lattice is proportional to vorticity of
corresponding plaquette in original lattice \cite{TJ,TJ1,TJ2}.
This shows that the zero vorticity condition in CCG model means
that the system is neutral.

This condition was applied before for the description of
numerical results on the square lattice \cite{TJ,TJ1}. In here we
apply this condition for lattices with two types of plaquettes,
 corresponding to CCG models with four types of charge. Our
result is in agreement with those without this explicit condition
and is an improvement on them \cite{unpub}.

Behavior of $E(f)$ is in close relation with behavior of critical
temperature $T_C(f)$ of lattice. In fact, when the ground state
energy is increased, we expect that the critical temperature is
lowered and vice versa \cite{lobb} . Our result is in agreement
with the mean field approach for $T_C(f)$\cite{LMF}.

This model can be mapped onto the frustrated $XY$
model\cite{k2002,lobb,CD}, which is the $XY$ model with unitary
coupling which is dependent on a parameter equivalent to the
frustration factor. This model is in close relation to the
Frenkel-Kontorova model\cite{R10,R101} which describes the
behavior of the ground state for a system of particles under the
effect of the lattice potential.

 The structure of the paper is
as follows: in section II, we introduce our model with the zero
vorticity condition; in section III we compare the results of the
model with the numerical study, comparing energies and
corresponding Fourier spectra; the final section is devoted to
Conclusions.
\section{The Model}
For a Josephson junction lattice, the Hamiltonian is,
\begin{equation}
H=-\sum_{<i,j>}cos(\gamma_{ij})\label{hamil}
\end{equation}
where $<i,j>$ means that $i$th node and $j$th node are connected
through a bond of lattice. $\gamma_{i,j}$ is the gauge invariant
phase difference. It is equal to
\begin{equation}
\gamma_{i,j}=\theta_i-\theta_j-A_{i,j}
\end{equation}
where $\theta$ is the superconducting phase and $A_{i,j}$ is the
integral of the vector potential from $i$th node to $j$th node.
We define the frustration factor as the ratio of magnetic flux
through a plaquette with unit area to the flux quantum.

The gauge invariant phase difference satisfies the fluxoid
quantization\cite{Tinkham},
\begin{equation}\label{zerovort}
V_k=\sum \gamma_{ij} =2\pi(n_k-a_kf)
\end{equation}
where $n_k$ is an integer and $a_k$ is the area of $k$th
plaquette. Summation is over the edges of $k$th plaquette. $V_k$
is vorticity of plaquette, and in transformation to CCG,
 is proportional to charges on $k$th site of the dual
lattice point\cite{TJ,TJ1,TJ2}. $n_k$ can have two values:
$\lfloor a_kf\rfloor$ or $\lfloor a_kf \rfloor+1$ where $\lfloor x
\rfloor$ means greatest integer less than $x$. It means that if
we have $l$ types of plaquettes, then we have $2l$ different
values for $V_k$ and therefore $2l$ types of charges in the
equivalent CCG model. We want to minimize this Hamiltonian with
respect to the superconducting phases. We use the Monte-Carlo (MC)
method\cite{MCm,MCm1} for the numerical minimization. We denote
this minimum by $E$ which is normalized by the number of
junctions.

From symmetry of a single plaquette, we can say that all four
gauge invariant phase differences are equal, and we denote it by
$\gamma$. Then from Eq. \ref{zerovort}, we can find
$\gamma=\pi/2(n-af)$, for two choices of $n$, and giving the
ground state energy of the plaquette as \cite{k2002},
\begin{eqnarray}\label{onepla}
 E_i=-cos(\pi/2(\lfloor af\rfloor+i-af))
\end{eqnarray}
for $i=0,1$. Now, we approximate the Hamiltonian as follows:
supposing that $k$th plaquette has vorticity $\lfloor
a_kf\rfloor+i_k$, then it has energy $E_{i_k}^k$ in IPM
approximation, and we can write Hamiltonian \ref{hamil} as sum of
these energies in IPM approximation:
\begin{equation}
H_{IPM}=\sum_{k}E_{i_k}^k
\end{equation}
where $i_k$ determines vorticity of $k$th plaquette. Now we
minimize above Hamilitonian with respect to $i_k$. In original
IPM, we choose $i_k$, such that $\lfloor a_kf\rfloor+i_k$ becomes
nearest integer to $a_kf$\cite{k2002}. Because of interaction
between plaquettes this is not exact, and some of the plaquettes
give the furthest integer as their vorticity. We suppose that this
deviation from IPM is such that total vorticity of lattice
becomes as small as possible. This is the zero vorticity
condition, with the total vorticity given as,
\begin{equation}
V_{total}=\sum_kV_k=2\pi(n_{total}-a_{lattice}f)
\end{equation}
where $n_{total}=\sum_kn_k$ and $a_{lattice}$ is the area of the
whole lattice. The condition for zero total vorticity reads,
\begin{equation}
V_{total}/N=\sum_kV_k/N\approx 0 \label{zerocir}
\end{equation}
where $N$ is total number of plaquettes.

This condition can be understood if we look at the relation
between Josephson junction lattice and CCG. It can be shown that
Josephson junction hamiltonian can be mapped to CCG on the dual
lattice, and in this transformation $V_k$ is proportional to
charge on the equivalent point in the dual lattice. Hence, the
 zero vorticity condition is
equivalent to the charge neutrality of CCG system.
\cite{TJ,TJ1,TJ2}

 Suppose that we have $k$ types of plaquettes in a lattice,
then each type can take on two energies, either $E_0$ or $E_1$,
bringing in $2k$ variables. But the total number of each type of
plaquette is determined and therefore we have $k$ constraints on
these variables and there are $k$ independent variables. Using
the above equation lets us remove one of these variables and
finally we have $k-1$ variables. If we apply the IPM
approximation, the Hamiltonian becomes linear in terms of these
variables and easily can be minimized. We denote this model as
IPM0.

We use the power spectrum of $E(f)$ to characterize it,
\begin{equation}
S(\omega)=\check{E}\check{E}^{*}
\end{equation}
where $\check{E}^{*}$ is the complex conjugate of $\check{E}$,
the Fourier transform of $E(f)$,
\begin{equation}
\check{E}(\omega)=\frac{1}{2\pi}\int_{-\infty}^{\infty}E(f)exp(2\pi
i\omega f)df
\end{equation}
The power spectrum was calculated in MATLAB by the $FFT$
method\cite{FFT}.

Now we focus on lattices with two types of plaquettes. We choose
two lattices: the Fibonacci ladder and the Penrose lattice. These
two cases are in close relation with each other. Ratio of areas
of plaquettes in both are the same and both make use of the
Fibonacci order. Therefore we can compare their energies to give
some insight about the difference between them.
 From the above discussion, we can find that these
Josephson junction lattices are equivalent to the $CCG$ model with
four types of charge.
\section{Lattice with two types of plaquettes}
In here we consider the above formulation for lattices with two
types of plaquettes which we denote them by $L$ and $S$. $S$
plaquette has area $1$ and $L$ plaquette has area $\alpha$.
Density of $S$ plaquettes is $N_S$ and that of the $L$ plaquettes
is $N_L$, therefore $N_L+N_S=1$. We suppose that the density of
$S$ ($L$) plaquettes with
 energy $E_0$ is $n_S$($n_L$), therefore from Eq.
\ref{zerocir}, we can find that
\begin{equation}
n_S+n_L=1+N_L(\lfloor \alpha f\rfloor-\alpha f)+N_S(\lfloor
f\rfloor- f)\label{eq1}
\end{equation}

This condition for a lattice with one type of plaquette
($a_L=a_S=1$, $n_S=n_L=n$) is reduced to
\begin{equation}
n=1+(\lfloor f\rfloor - f)
\end{equation}
which is in agreement with the previous result \cite{TJ,TJ1}. The
Hamiltonian for a lattice with two types of plaquettes within the
IPM is
\begin{equation}
H_{IPM}=(n_SE_0^S+(N_S-n_S)E_1^S+n_LE_0^L+(N_L-n_L)E_1^L)
\end{equation}
Now we minimize Hamiltonian with respect to these two densities
with condition Eq. \ref{eq1}. If we increase $n_L$ by one
(therefore decrease $n_L$ by one), then change in Hamiltonian is
$\Delta E=E_0^L-E_1^L-E_0^S+E_1^S$, therefore if $\Delta E>0$ then
$n_s=min(N_S,q(f))$ and if $\Delta E<0$ then $n_L=min(N_L,q(f))$
where $min(x,y)$ means the minimum between $x$ and $y$ . We can
write these results as
\begin{eqnarray}
n_S=N_S\theta(\Delta E)+(q(f)-N_L)(1-\theta(\Delta E))\\
 n_L=N_L(1-\theta(\Delta E))+(q(f)-N_S)\theta(\Delta E)
\end{eqnarray}
where $\theta(x)$ is the step function which is zero for negative
$x$ and one for positive $x$, and $q(f)$ represents the right
hand side of Eq. \ref{eq1}.

Now we apply these results to a ladder with two types of
plaquettes and the 2D Penrose lattice.
\subsection{ Fibonacci Ladder}
\begin{figure}[ht!]
\centerline{\includegraphics[width=10cm]{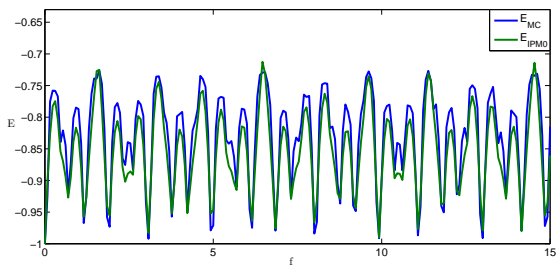}}
\vspace{0.1cm}
 \caption{Numerical and IPM0 energy for the Fibonacci ladder }
\vspace{-0.3cm} \label{etau}
\end{figure}

\begin{figure}[ht!]
\centerline{\includegraphics[width=10cm]{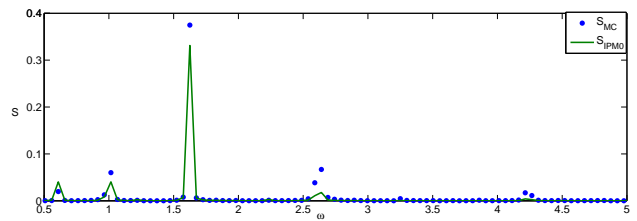}}
\vspace{0.1cm}
 \caption{Numerical and IPM0 Fourier spectrum for the Fibonacci ladder }
\vspace{-0.5cm} \label{ftau}
\end{figure}

For this lattice we have, $\alpha=\tau=(1+\sqrt(5))/2$ and the
plaquettes appear in the Fibonacci order, defined as follows.
Denoting the structure in its $m$th step of construction by
$U_{m}$, the structure is constructed recursively by the rule
$U_{m+2}=U_{m+1}+U_{m}$, where summation means adjacent placement:
to get the structure in step $(m+2)$, put the structure in step
$(m+1)$, first and that of step $m$, to its right. This generates
the sequence
\begin{equation}
S,L,LS,LSL,LSLLS,LSLLSLSL,...
\end{equation}
In the limit of an infinite structure this defines a
quasicrystalline lattice such that ratio of $N_L$ to $N_S$ is
$\tau$. The energy and Fourier spectra for MC and IPM0 are given
in Figs. \ref{etau} and \ref{ftau}.

As we see, the agreement between two results is very good. From
Fourier spectrum comparison, we see that there are peaks on
$\omega=1,\tau,1+\tau,\tau-1$, for both models. This agrees with
the results of IPM without the zero vorticity
condition\cite{unpub}.

We also compare results of these two methods for other types of
lattices with different ratio of areas and again find good
agreement.

\subsection{Two dimensional example: The Penrose lattice}
Penrose lattice is an example of 2D quasicrystals \cite{Penrose}
which has two types of rhombi with areas $sin(2\pi/5)$ and
$sin(\pi/5)$. It is known that minimization of Josephson Junction
Hamiltonian on this lattice, as well as on the Fibonacci ladder,
does not result in a periodic $E(f)$; similarly, $T_C(f)$ is
aperiodic \cite{R13,R131,R132,R133}. Our numerical results agree
with this observation.

\begin{figure}[ht!]
\centerline{\includegraphics[width=10cm]{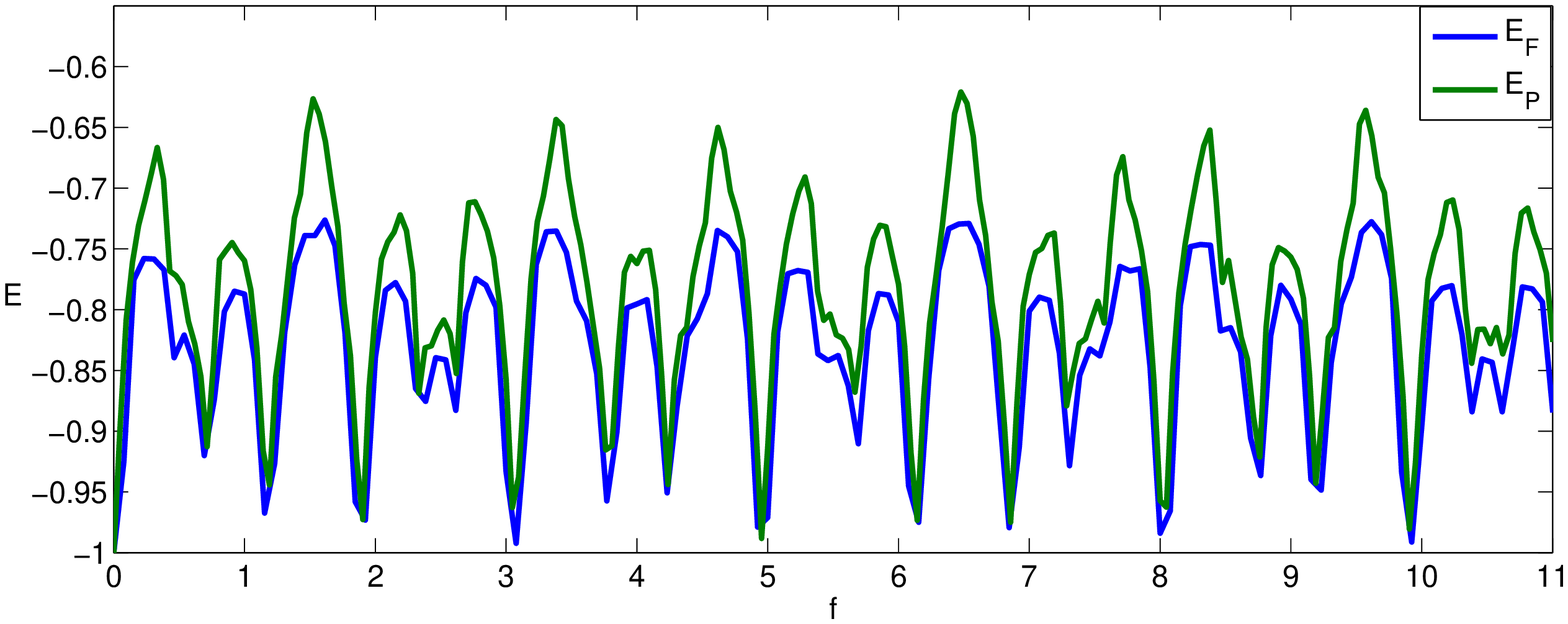}}
\vspace{0.5cm}
 \caption{Comparison between energy for Penrose lattice $E_P$ and Fibonacci ladder $E_F$. }
\vspace{-0.5cm} \label{efp}
\end{figure}

We compare the ground state energy for this lattice with its
one-dimensional counterpart, the Fibonacci ladder. This shown in
Fig. \ref{efp}.
We see that the ground state energy for the Penrose lattice is
greater than that of the Fibonacci ladder. This can be related to
the constraints involved: in the two dimensional case there are
more neighboring plaquettes than in the ladder geometry.

\begin{figure}[ht!]
\centerline{\includegraphics[width=10cm]{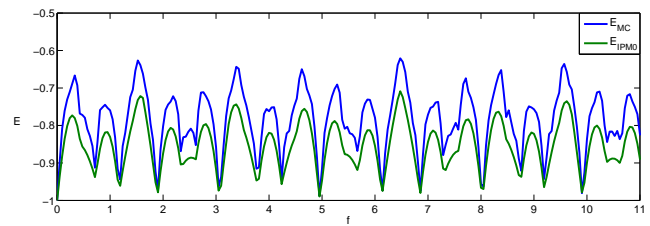}}
\vspace{0.1cm}
 \caption{Numerical and IPM0 energy for Penrose lattice }
\vspace{-0.5cm} \label{epenrose}
\end{figure}

\begin{figure}[ht!]
\centerline{\includegraphics[width=10cm]{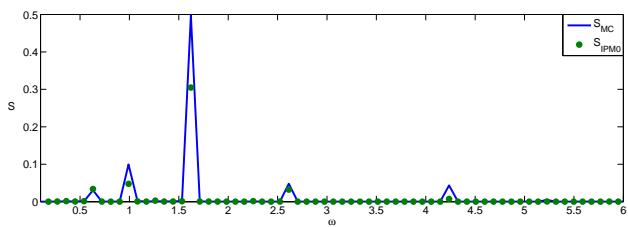}}
\vspace{0.1cm}
 \caption{Numerical and IPM0 Fourier spectrum for Penrose lattice }
\vspace{-0.5cm} \label{fpenrose}
\end{figure}

The ground state energy using MC and IPM0 and their corresponding
Fourier spectra for the Penrose lattice are shown in Figs.
\ref{epenrose} and \ref{fpenrose}.

We can see that the results of IPM0 are in agreement generally
with Monte-Carlo results. Also IPM0 predicts the main frequencies
of $E(f)$ correctly.

\section{Conclusion}
We enter a condition in the calculation of the ground state
energy of Josephson junction lattices, namely that the total
vorticity of lattice be zero. This condition along with the
Independent Plaquette Model (IPM) gives us the properties of the
ground state energy with good approximation. It can describe the
main peaks of the Fourier spectrum of $E(f)$ on areas of
plaquettes and their sum and difference. In 2D the accuracy of
the model is not as good as in the case of the ladder, because of
greater effects of neighboring plaquettes: In 2D a plaquette has
more neighbors than in the ladder, and therefore IPM is worse in
2D than in the effectively 1D ladder.

Zero vorticity condition can be understood when we look at the
connection of Josephson junction Hamiltonian with CCG. In CCG
model vorticity of plaquette plays the role of charge on its
corresponding dual lattice site. Therefore zero vorticity
condition in CCG means that system of charge is neutral, which
naturally leads to a lower energy for the ground state.

\section{Acknowledgments}

I thank the Institute for Advanced Studies in Basic Sciences for
supporting this research. I also thank M. R. Kolahchi for useful
comments on the manuscript.

\end{document}